\documentstyle[epsfig]{mn}

\begin{document}

\title[Structural evolution in ellipticals]{Structural evolution in
elliptical galaxies: the age -- shape relation}

\author[Ryden, Forbes \& Terlevich]{
 Barbara S. Ryden$^1$\thanks{email: ryden@astronomy.ohio-state.edu},
 Duncan A. Forbes$^2$ and A.I.~Terlevich$^3$\\
 $^1$Department of Astronomy, The Ohio State University, Columbus OH 43210 USA\\
 $^2$Centre for Astrophysics \& Supercomputing, Swinburne University, Hawthorn VIC 3122,
 Australia\\
 $^3$School of Physics and Astronomy, University of Birmingham, Birmingham B15 2TT}

\date{Received:\ \ \ Accepted: }
 
\pagerange{\pageref{firstpage}--\pageref{lastpage}}%

\maketitle

\label{firstpage}

\begin{abstract}
We test the hypothesis that the apparent axis ratio of an
elliptical galaxy is correlated with the age of its stellar
population. We find that old ellipticals (with
estimated ages $t > 7.5 {\rm\,Gyr}$)
are rounder on average than younger ellipticals. The
statistical significance of this shape difference is greatest
at small radii; a Kolmogorov--Smirnov test comparing the axis
ratios of the two populations at $R = R_e/16$ yields
a statistical significance greater than 99.96\%.
The relation between age and apparent shape is linked to
the core/power-law surface brightness profile dichotomy.
Core ellipticals have older stellar populations, on average,
than power-law ellipticals and are rounder in their
inner regions. Our findings are consistent with a scenario
in which power-law ellipticals are formed in gas-rich mergers,
while core ellipticals form in dissipationless mergers, with
cores formed and maintained by the influence of a binary
black hole.

\end{abstract}


\section{Introduction}

In the standard classification scheme devised by Hubble (1926),
the apparent shape of an elliptical galaxy is designated by
a single number; each elliptical is given the label `E{\it n}',
where {\it n} is ten times the ellipticity of the galaxy's projected image,
rounded to the nearest integer. In actuality, of course, the
structure of an elliptical galaxy is too complicated to be
summed up in a single number. When the surface brightness
of an elliptical is fitted with elliptical isophotes of
semimajor axis $a$ and semiminor axis $b$, it is found
that the axis ratio $q \equiv b/a$ is a function of the
isophotal radius $R \equiv (ab)^{1/2}$. The majority of
ellipticals have axis ratios $q$ that decrease with $R$ or
remain roughly constant; however, some ellipticals become
rounder with increasing $R$, or have $q(R)$ that
varies irregularly (Bettoni et al. 1997).

It isn't surprising that the shapes of elliptical galaxies
can vary with radius, since the processes that sculpt
galaxies are different in the inner and outer regions.
It appears from a growing body of evidence that most,
if not all, elliptical galaxies harbor a central supermassive
black hole (Kormendy \& Richstone 1995). A black hole
can have a significant effect on the stellar distribution in
an elliptical galaxy's inner regions; by disrupting box
orbits, it drives an initially triaxial galaxy to
a more nearly axisymmetric shape (Gerhard \&
Binney 1985; Norman et al. 1985; Valluri \& Merritt 1998).
The outer regions of elliptical galaxies, particularly in
rich clusters and compact groups, will be shaped by tidal
encounters with neighboring galaxies.

At a given time, we expect the shape of an elliptical galaxy
to be a function of radius. Moreover, at a given radius within
a galaxy, we expect the shape to be a function of time.
In the hierarchical clustering model for the formation of
structure, ellipticals form by the merger of smaller galaxies.
Mergers of galaxies with roughly equal mass create merger
remnants which are flattened; the ratio of shortest
to longest axis is typically $\gamma \equiv c/a \sim 0.5$
(Barnes 1992; Springel 2000). For a dissipationless merger,
the elliptical remnant may be oblate, prolate, or triaxial,
depending on the initial geometry of the merger (Barnes 1992).
The presence of gas in the merging galaxies doesn't strongly
affect the axis ratio $\gamma$, but tends to make the
remnant more nearly oblate (Springel 2000). Thus, the
overall shape of an elliptical galaxy depends on whether
the most recent major merger in its history was gas-rich
or gas-poor. In addition, if each of the two merging progenitors
contains a central black hole, the two black holes will
significantly affect the structure of the merger remnant
as they gradually spiral together and coalesce. By ejecting
stars, the black hole binary carves out a central core
with stellar density $\rho \propto r^{-\alpha}$, with $\alpha
\sim 0.5$ (Ebisuzaki et al. 1991; Makino \& Ebisuzaki 1996).
The binary black hole undergoes a random walk as it ejects
stars from the core (Merritt 2001),
tending to create a cuspy core which is more nearly spherical
than the flattened, $\gamma \sim 0.5$, merger remnant. 

In view of these physical considerations, we expect the shape
of an elliptical galaxy to depend on its merger history and
to be modified in its central regions by the effects of
massive black hole binaries. In general, elliptical galaxies formed
in a major merger will become rounder with time (at least
in their inner regions). The same physical processes that affect
the shape of an elliptical galaxy may also modify its luminosity
profile; thus, we might expect `core' ellipticals
to differ significantly in shape from `power-law' ellipticals.
The purpose of this paper is to test these na{\"\i}ve expectations
by examining the projected axis ratios $q$ of a sample of nearby
bright elliptical galaxies, as a function of galaxy assembly age ($t$), isophotal
radius ($R$), and luminosity profile type (core or power-law).

In section 2 of this paper, we describe how we assign an age $t$
to each galaxy in our sample. In section 3, we describe how
we determine the shape profile $q(R)$ for each galaxy. In section
4, we examine the relation between $q$ and $t$ at different
fiducial radii. In section 5, we discuss how the age -- shape
relation differs for core ellipticals and for power-law ellipticals.
Finally, in section 6, we consider how these results shed
light on the evolution of elliptical galaxies and the origin
of the core/power-law dichotomy.

\section{Age determination}

If an elliptical galaxy is formed by the successive merger of
smaller galaxies, then assigning an age to that galaxy
becomes an exercise in ambiguity. In this paper, we are
concerned with how the shape of a galaxy changes as it
evolves dynamically. Thus, the most useful definition of
a galaxy's age, for our purposes, is the time that has elapsed
since the galaxy last underwent a major merger. 
If the merging galaxies contain gas, then the gas loses angular momentum
during the course of the merger and flows to the centre, where it
triggers a brief but intense burst of star formation (Barnes \&
Hernquist 1991, 1996; Mihos \& Hernquist 1994, 1996). The
time since the last major merger (if the merging progenitors
contained significant amounts of gas) should be equal to
the age of the youngest stars in the central region of the
elliptical.

The age determinations used in this paper are drawn from
the recent catalogue of Terlevich \& Forbes (2000). This
catalogue contains galaxies for which there exist high-quality
H$\beta$ and [MgFe] absorption line indices. Using the stellar
population model of Worthey (1994), these line indices are
used to break the age/metallicity degeneracy, giving separate
age and metallicity estimates for the galaxies in the catalogue.
The line indices used by Terlevich \& Forbes (2000) come
from the galaxies' central regions and are luminosity weighted.
Thus, they are dominated by the last major burst of star formation
at the centre of each galaxy. The age determinations are therefore
likely to reflect the time since the most recent major
merger in the galaxy's history. (One caveat must be added:
if a major merger is purely dissipationless, it will not
trigger a burst of star formation, and thus will not leave
its mark on the stellar population.)

Terlevich \& Forbes (2000) provide ages
for $\sim 150$ relatively nearby galaxies. We eliminated from
our sample galaxies with estimated ages $t > 17 {\rm\,Gyr}$;
these galaxies may have authentically old stellar populations,
but they may also be suffering from residual H$\beta$ emission.
Of the remaining galaxies, we select the 68 galaxies classified
by Terlevich \& Forbes as E galaxies, the 5 galaxies classified
as cD galaxies (NGC 1399, NGC 2832, NGC 4839, NGC 4874, and IC 5358),
and the one galaxy classified as cE (NGC 221); the morphological
types used by Terlevich \& Forbes were taken from the NASA/IPAC
Extragalactic Database.
The 74 galaxies in our sample are listed in Table 1, along with their
ages as estimated by Terlevich \& Forbes (2000). It is important
to note that the overall calibration of the ages is somewhat uncertain,
although the relative ranking of the ages is quite robust.

\begin{table*}
\begin{minipage}{175mm}
\caption{Elliptical galaxy ages and axis ratios}
\begin{tabular}{lccccccccccc}
Galaxy &  Age  & Profile  &  $R_e$   & $q(R_e/16)$  & $q(R_e/8)$ &
$q(R_e/4)$ & $q(R_e/2)$ & $q(R_e)$  &   $q(2 R_e)$   & \multicolumn{2}{c} {Isophotal data} \\
 name  & [Gyr] & type$^a$ & [arcsec] &  &  &
 &  &  &  & HST$^b$ & Ground-based$^c$ \\ \hline
NGC221  & 3.8 & $\backslash$ & 38.6 & 0.71 & 0.72 & 0.73 & 0.77 & 0.83 & 0.85 & a & 1 \\
NGC315  & 4.9 & - & 58.5 & 0.73 & 0.71 & 0.72 & 0.74 & 0.77 &  -    & b & 2,3 \\
NGC547  & 7.6 & $\cap$ & 12.6 &  -    & 0.98 & 0.94 & 0.89 & 0.83 &  -    & z & 4 \\
NGC584  & 2.1 & $\backslash$ & 27.4 & 0.75 & 0.80 & 0.80 & 0.69 & 0.65 &  -    & z & 4 \\
NGC636  & 3.6 & - & 18.9 &  -    &  -    & 0.92 & 0.90 & 0.84 & 0.82 & - & 4,5 \\
NGC720  & 3.4 & $\cap$ & 39.6 & 0.81 & 0.72 & 0.64 & 0.58 & 0.55 & 0.55 & c & 3,4,6,7,8,9 \\
NGC821  & 7.2 & $\backslash$ & 45.4 & 0.61 & 0.61 & 0.63 & 0.64 & 0.72 &  -    & d & 4,9 \\
NGC1209 & 15. & - & 17.7 &  -    &  -    &  -    & 0.56 & 0.44 & 0.43 & - & 7 \\
NGC1339 & 7.5 & $\cap$ & 16.9 & 0.85 & 0.83 &  -   & 0.76 & 0.70 & 0.71 & z & 10 \\
NGC1373 & 8.9 & $\cap$ & 11.8 & 0.75 & 0.77 & 0.79 & 0.77 & 0.79 & 0.93 & z & 10 \\
NGC1374 & 9.8 & $\cap$ & 30.0 & 0.90 & 0.87 & 0.90 & 0.88 & 0.91 & 0.93 & z & 10 \\
NGC1379 & 7.8 & $\cap$ & 42.4 & 0.99 & 0.98 & 0.98 & 0.98 & 0.96 & 0.98 & z & 5,10 \\
NGC1399 & 5.0 & $\cap$ & 42.4 & 0.90 & 0.88 & 0.87 & 0.89 & 0.91 & 0.96 & c & 5,8,9,10 \\
NGC1404 & 5.0 & $\cap$ & 26.7 & 0.85 & 0.84 & 0.85 & 0.88 & 0.88 & 0.86 & z & 5,9,10 \\
NGC1419 & 8.2 & $\backslash$ & 10.9 & 0.77 & 0.78 & 0.88 & 0.98 & 0.99 & 0.96 & z & 10 \\
NGC1427 & 6.5 & $\backslash$ & 32.9 & 0.68 & 0.74 & 0.72 & 0.71 & 0.68 & 0.71 & e,f & 9,10 \\
NGC1453 & 7.6 & - & 28.0 &  -    &  -    & 0.85 & 0.84 & 0.84 &  -    & - & 4,8 \\
NGC1549 & 7.6 & - & 47.6 &  -    & 0.84 & 0.85 & 0.87 & 0.88 & 0.92 & - & 5,9,11 \\
NGC1600 & 6.9 & $\cap$ & 47.6 & 0.76 & 0.66 & 0.65 & 0.68 &  -    &  -    & d,g & 3,4 \\
NGC1700 & 2.3 & $\backslash$ & 13.7 & 0.74 & 0.74 & 0.66 & 0.73 & 0.74 & 0.71 & c,f & 4,5,7,9 \\
NGC2778 & 8.2 & $\backslash$ & 16.5 &  -   &  -   &  -   & 0.79 & 0.76 & 0.84 & - & 3 \\
NGC2832 & 12. & $\cap$ & 25.5 & 0.83 & 0.79 & 0.80 & 0.73 & 0.71 & 0.61 & c & 3,12 \\
NGC2865 & 1.5 & $\cap$ & 11.7 & 0.70 & 0.71 & 0.83 & 0.80 & 0.72 & 0.75 & z & 11,13 \\
NGC3078 & 14. & - & 23.8 &  -    &  -    &  -    &  -    &  -    &  -    & - & - \\
NGC3377 & 4.1 & $\backslash$ & 33.7 & 0.42 & 0.42 & 0.53 & 0.50 & 0.55 & 0.63 & c & 3,9,11 \\
NGC3379 & 9.3 & $\cap$ & 35.2 & 0.89 & 0.90 & 0.92 & 0.91 & 0.87 & 0.87 & g & 3,4,9,11 \\
NGC3585 & 3.1 & $\cap$ & 39.6 & 0.65 & 0.56 & 0.50 &  -    &  -    &  -    & z & - \\
NGC3605 & 5.8 & $\backslash$ & 17.3 & 0.66 & 0.67 & 0.66 & 0.64 & 0.59 & 0.66 & c & 3,11 \\
NGC3608 & 10. & $\cap$ & 35.2 & 0.84 & 0.80 & 0.82 & 0.79 & 0.77 & 0.78 & c,f & 9,11 \\
NGC3818 & 5.0 & - & 21.2 &  -    &  -    &  -    & 0.60 & 0.64 & 0.67 & - & 11 \\
NGC4073 & 7.5 & $\cap$ & 55.9 & 0.72 &  -    &  -    &  -    &  -    &  -    & z & - \\
NGC4239 & 5.5 & $\backslash$ & 16.1 & 0.57 & 0.59 & 0.49 &  -    &  -    &  -    & c & - \\
NGC4261 & 9.4 & $\cap$ & 38.6 & 0.75 & 0.73 & 0.75 & 0.81 & 0.84 & 0.85 & b,d & 3,4,9,10 \\
NGC4278 & 8.4 & $\cap$ & 32.9 & 0.84 & 0.85 & 0.85 & 0.87 & 0.91 & 0.91 & f & 3,9 \\
NGC4339 & 7.9 & $\backslash$ & 30.7 & 0.98 & 0.96 & 0.95 & 0.93 & 0.93 & 0.93 & z & 10 \\
NGC4374 & 11. & $\cap$ & 54.6 & 0.83 & 0.81 & 0.85 & 0.87 & 0.91 & 0.96 & b,d & 3,4,9,11,14 \\
NGC4387 & 13. & $\backslash$ & 15.4 & 0.80 & 0.79 & 0.68 & 0.65 & 0.59 & 0.64 & c & 3,11,14 \\
NGC4458 & 16. & $\backslash$ & 26.7 & 0.76 & 0.73 & 0.88 & 0.89 & 0.92 & 0.93 & c & 11,14 \\
NGC4472 & 8.5 & $\cap$ & 104. & 0.91 & 0.85 & 0.83 & 0.83 & 0.83 & 0.77 & g & 3,4,10 \\
NGC4478 & 4.1 & $\backslash$ & 14.0 & 0.62 & 0.79 & 0.83 & 0.80 & 0.82 & 0.84 & h & 3,11,14 \\
NGC4489 & 2.6 & - & 32.2 &  -    &  -    & 0.92 & 0.92 & 0.91 & 0.94 & - & 11 \\
NGC4551 & 5.2 & $\backslash$ & 17.7 & 0.62 & 0.71 & 0.74 & 0.74 & 0.71 & 0.76 & c & 3,11,14 \\
NGC4552 & 9.6 & $\cap$ & 30.0 & 0.94 & 0.94 & 0.96 & 0.95 & 0.88 & 0.90 & f,g & 4,14 \\
NGC4564 & 5.9 & $\backslash$ & 21.7 & 0.75 & 0.74 & 0.65 & 0.47 & 0.42 & 0.47 & h & 9,14 \\
NGC4649 & 11. & $\cap$ & 73.6 & 0.91 & 0.86 & 0.84 & 0.80 & 0.80 & 0.79 & g & 3,4,14 \\
NGC4697 & 8.2 & $\backslash$ & 75.4 & 0.58 & 0.61 & 0.56 & 0.58 & 0.66 & 0.64 & c & 3,6,9,11 \\
NGC4839 & 15. & - & 28.6 &  -    &  -    & 0.76 & 0.68 & 0.58 &  -    & - & 13 \\
NGC4860 & 12. & $\cap$ &  8.6 & 0.92 & 0.90 & 0.84 &  -    &  -    &  -    & z & - \\
NGC4869 & 15. & - &  8.5 &  -    &  -    &  -    &  -    & 0.90 & 0.96 & - & 2,13 \\
NGC4874 & 13. & $\cap$ & 61.2 & 0.90 & 0.94 & 0.96 & 0.87 & 0.90 &  -    & c & 3,13 \\
NGC4876 & 2.1 & $\cap$ &  6.0 &  -    &  -    & 0.95 &  -    & 0.70 & 0.70 & z & 13 \\
NGC4908 & 12. & - &  9.5 &  -    &  -    &  -    &  -    &  -    &  -    & - & - \\
NGC4926 & 13. & $\cap$ & 11.4 & 0.80 & 0.79 & 0.80 & 0.84 & 0.88 &  -    & z & 13 \\
NGC4952 & 6.6 & - &  9.0 &  -    &  -    &  -    &  -    & 0.72 & 0.69 & - & 15 \\
NGC4957 & 4.9 & $\backslash$ & 14.7 & 0.78 & 0.76 & 0.76 & 0.78 & 0.75 & 0.76 & z & 15 \\
NGC5018 & 1.5 & - & 25.0 &  -    &  -    & 0.65 & 0.67 & 0.72 & 0.75 & - & 9 \\
NGC5638 & 7.0 & - & 28.0 &  -    &  -    & 0.95 & 0.91 & 0.89 & 0.88 & - & 3,11 \\
NGC5812 & 5.0 & $\cap$ & 23.8 & 0.99 & 0.98 & 0.97 & 0.96 & 0.93 & 0.90 & z & 8 \\
NGC5831 & 2.6 & $\backslash$ & 26.7 & 0.73 & 0.69 & 0.71 & 0.84 & 0.91 & 0.87 & z & 3,11 \\
NGC5846 & 12. & $\cap$ & 82.6 & 0.92 & 0.91 & 0.92 &  -    &  -    &  -    & z & 4 \\
NGC6127 & 9.3 & - & 21.7 &  -    &  -    &  -    &  -    &  -    &  -    & - & - \\
NGC6702 & 1.9 & - & 28.6 &  -    &  -    & 0.72  & 0.79 &  -   &   -   & - & 4 \\
NGC6958 & 12. & - & 20.8 &  -    &  -    &  -    & 0.86 & 0.85 &   -   & - & 8 \\
\end{tabular}
\end{minipage}
\end{table*}

\begin{table*}
\begin{minipage}{175mm}
\contcaption{}
\begin{tabular}{lccccccccccc}
Galaxy &  Age  & Profile &   $R_e$  & $q(R_e/16)$ & $q(R_e/8)$ &
$q(R_e/4)$ & $q(R_e/2)$ & $q(R_e)$ & $q(2 R_e)$ & \multicolumn{2}{c} {Isophotal data} \\
       & [Gyr] &   type$^a$ & [arcsec] &             &            &
           &            &          &            & HST$^b$ & Ground-based$^c$ \\ \hline
NGC7052 & 11. & $\cap$ & 20.8 & 0.80 & 0.70 & 0.68 & 0.55 & 0.48 &  -    & b,d & 2,4 \\
NGC7454 & 5.2 & - & 24.4 &  -    &  -    & 0.70 & 0.64 & 0.69 &  -    & - & 4 \\
NGC7562 & 11. & $\backslash$ & 23.8 & 0.81 & 0.81 & 0.80 & 0.72 & 0.68 &  -    & z & 4,8 \\
NGC7619 & 9.0 & $\cap$ & 32.2 & 0.77 & 0.74 & 0.73 & 0.77 & 0.81 & 0.83 & z & 4,5 \\
NGC7626 & 12. & $\cap$ & 37.8 & 0.90 & 0.90 & 0.89 & 0.87 & 0.83 &  -    & b,d,e,f & 2,3,4,5,8 \\
NGC7785 & 8.3 & $\backslash$ & 26.7 & 0.64 & 0.64 & 0.63 & 0.58 & 0.50 &  -    & z & 4,8 \\
IC2006  & 6.0 & - & 28.6 &  -    &  -    & 0.90 & 0.89 & 0.88 & 0.88 & - & 16 \\
IC4045  & 14. & $\cap$ &  5.7 &  -    & 0.88 & 0.76 &  -    & 0.68 &  -    & z & 13 \\
IC4051  & 12. & $\backslash$ & 22.9 & 0.82 & 0.85 & 0.85 & 0.80 & 0.70 &  -    & z & 13 \\
IC5358  & 16. & - & 18.5 &  -    &  -    &  -    &  -    &  -    &  -    & - & - \\
E274G06 & 12. & - &  -   &  -    &  -    &  -    &  -    &  -    &  -    & - & - \\
\end{tabular}
\medskip
$^a$ $\backslash$ = Power-law, $\cap$ = Core
\par
$^b$ (a) Lauer et al. 1998, (b) Verdoes Kleijn et al. 1999, (c) Lauer et al. 1995,
(d) Quillen et al. 2000, (e) Forbes et al. 1995, (f) Carollo et al. 1997,
(g) Lauer et al. 2000, (h) van den Bosch et al. 1994, (z) this work.
\par
$^c$ (1) Peletier 1993, (2) de Juan et al. 1994, (3) Peletier et al. 1990,
(4) Lauer 1985, (5) Franx et al. 1989, (6) Jedrzejewski et al. 1987,
(7) Capaccioli et al. 1988, (8) Sparks et al. 1991, (9) Goudfrooij et al. 1994,
(10) Caon et al. 1994, (11) Jedrzejewski 1987, (12) Postman \& Lauer 1995,
(13) Jorgensen et al. 1992, (14) Caon et al. 1990, (15) Mehlert et al. 2000,
(16) Schweizer et al. 1989.
\end{minipage}
\end{table*}

\section{Apparent shape determination}

The 74 galaxies in our sample are relatively nearby, with distances
ranging from $0.72 {\rm\,Mpc}$ for NGC 221 to $112 {\rm\,Mpc}$ for
IC 5358, with a median distance of $23 {\rm\,Mpc}$
(assuming $H_0 = 75 {\rm\,km} {\rm\,s}^{-1} {\rm\,Mpc}^{-1}$).
The galaxies are a mix of field galaxies and galaxies from groups
and clusters, with 9 galaxies from the Fornax cluster, 11 from the
Virgo cluster, and 11 from the Coma cluster. Most of the galaxies
have photometric data available in the published literature.
Values of the effective radius $R_e$ for each galaxy in the
sample were taken from Faber et al. (1989), when available.
For those few galaxies not assigned a value of $R_e$ by Faber et al.
(1989), the effective radius was taken from other sources.
Table 1 gives the adopted value of $R_e$ for each galaxy.

Once a value of $R_e$ was assigned to each galaxy, we used
published isophotal data to find the axis ratio $q \equiv b/a$ at six
reference radii: $R \equiv (ab)^{1/2} = 2^n R_e$, where $n = -4$, $-3$,
$\dots$, $+1$. We found published isophotal fits based on
ground-based data for 65 galaxies in our sample, and fits
based on Hubble Space Telescope (HST) data for 29 galaxies
in our data.  We searched the HST archive for additional
images in which galaxies from our sample were located on
the PC chip of WFPC2. Galaxies with substantial central dust were
excluded from analysis, leaving 24 additional galaxies. These were
then modelled using the ISOPHOTE package in STSDAS.
The sources of isophotal
information for each galaxy are listed in Table 1. Fortunately,
the axis ratio $q$ is a robust parameter which does not significantly
depend on the isophote-fitting algorithm used. Moreover, for
elliptical galaxies, $q$ does not strongly depend on the filter used.
Thus, for galaxies with multiple sources of isophotal information, the
values of $q$ tabulated in Table 1 are simply the average value
of $q$ taken from all relevant sources. To minimize the effects of
seeing, we discarded all isophotes with $R$ less than 3 times the
FWHM of the relevant observation (assumed to be 0.1 arcsec for HST).

\begin{figure}
\centering \epsfig{figure=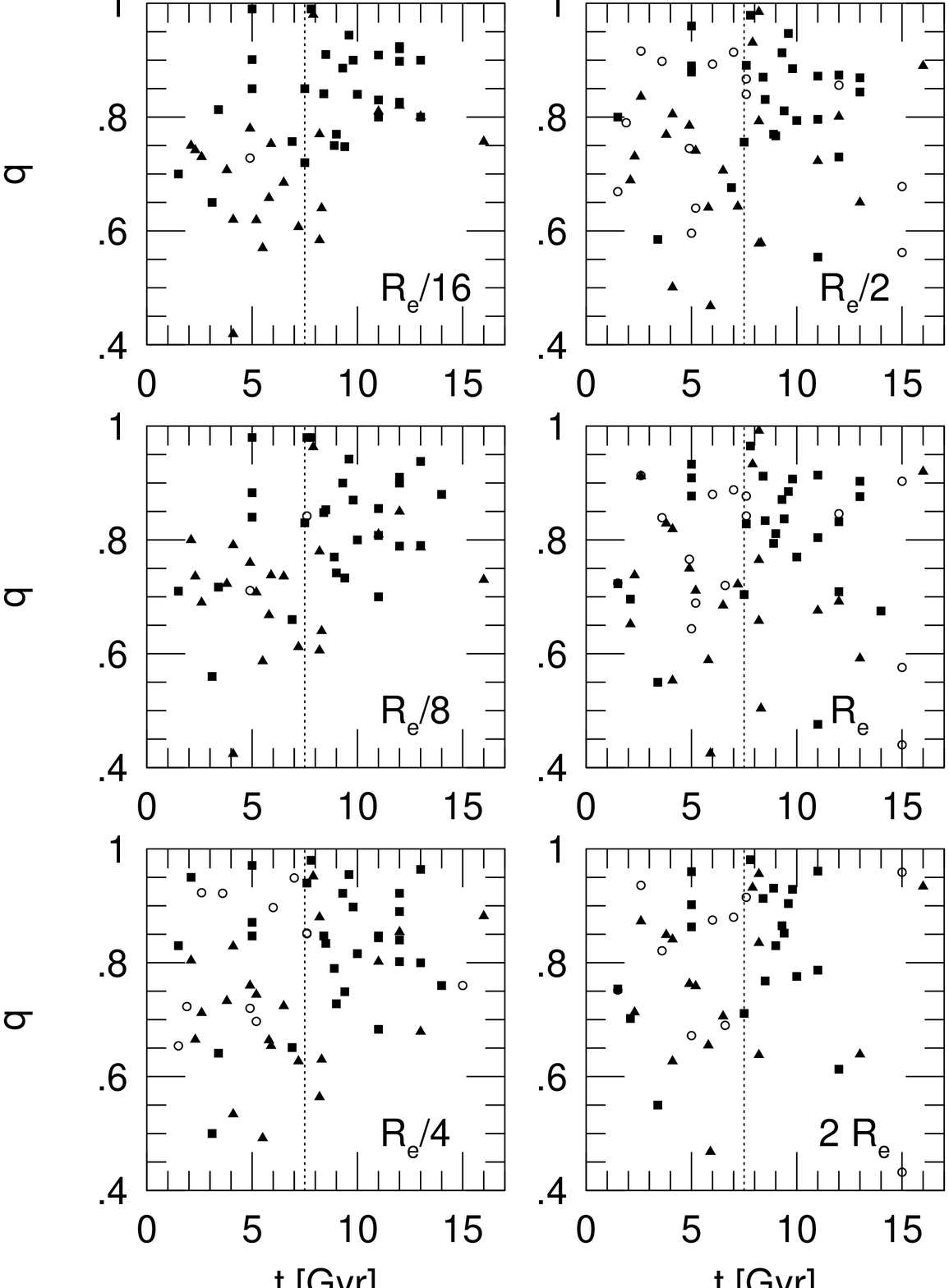, width=8.4cm}
\caption{
Isophotal axis ratio $q$ as a function of age estimate $t$ at
$R = R_e/16$, $R_e/8$, $R_e/4$, $R_e/2$, $R_e$, and $2 R_e$. Galaxies
with core profiles are indicated by squares, galaxies with power-law
profiles are indicated by triangles, and galaxies with unknown
profile type are indicated by open circles.}
\label{figure1}
\end{figure}

\section{The age -- shape relation}

Figure 1 shows a plot of the apparent axis ratio $q$ versus
the computed galactic age $t$ at our six reference radii,
$R_e/16$ through $2 R_e$.  In each panel, each point
represents a different galaxy. Particularly at the smallest
reference radius, $R_e/16$, there is a significant correlation
between $q$ and $t$, with old galaxies tending to
be rounder than young galaxies. The tightness of the
correlation decreases at larger radii.

To quantify the statistical significance of the difference
in shape between young and old ellipticals, we ran a number
of Kolmogorov--Smirnov tests, comparing the distribution
of $q$ for young galaxies ($t \leq t_0$) with that for
old galaxies ($t > t_0$). By computing the KS probability
$P_{\rm KS}$ for different values of the dividing time $t_0$,
we found $t_{0,m}$, the value of $t_0$ which minimizes
$P_{\rm KS}$, and hence maximizes the shape difference between
young and old ellipticals. Interestingly, for all six of
our reference radii, $P_{\rm KS}$ is minimized at $t_{0,m} \approx
7.5 {\rm\,Gyr}$. (At $R = R_e/16$, there is an additional
minimum in $P_{\rm KS}$, of comparable depth, at $t_0 = 9.5
{\rm\,Gyr}$). Since the sample of galaxies used to determine
$t_{0,m}$ is different at different values of $R$, though,
the prudent reader should not place too much emphasis on
the dependence of $t_{0,m}$; the more physically significant
relation is the dependence of $q$ on $t$ at a given value
of $R$.
For the rest of this paper, we will define `young' ellipticals
as having $t \leq 7.5 {\rm\,Gyr}$ and `old' ellipticals
as having $t > 7.5 {\rm\,Gyr}$. Table 2 gives the numbers
$N_{\rm young}$ and $N_{\rm old}$ of young and old ellipticals
in our sample at each reference radius, as well as the
mean axis ratios $\overline{q}_{\rm young}$ and $\overline{q}_{\rm old}$
and the KS probability measuring the difference in the shape distribution
for young and old galaxies.
Although the probability $P_{\rm KS} = 0.0034$ measured at $R_e/16$
is impressively small, remember that the dividing line $t_0 = 7.5
{\rm\,Gyr}$ between young and old galaxies was specifically chosen
to minimize $P_{\rm KS}$, and not set {\it a priori} from independent
considerations. To test the true statistical significance of the difference
in $q(R_e/16)$ between young and old galaxies, we did an analysis
involving bootstrap resampling. There are 50 galaxies in our sample for
which $q(R_e/16)$ was measured. Let $(t_1, t_2, \ldots, t_{50})$ be the
estimated ages of these galaxies and $(q_1, q_2, \ldots, q_{50})$ be their
values of $q(R_e/16)$. For each resampling of the data, we randomly drew
50 values of $t$, with replacement, from $(t_1, t_2, \ldots, t_{50})$
and paired them with 50 values of $q$ drawn randomly, with replacement,
from $(q_1, q_2, \ldots, q_{50})$. For these 50 random $(t,q)$ pairs,
we found the value of $t_0$ which minimizes the value of $P_{\rm KS}$
when comparing the values of $q$ for galaxies with $t>t_0$ to those of
galaxies with $t \leq t_0$. After performing $10^6$ resamplings, we found
that a fraction $f = 0.0012$ of the resampled data sets had a minimized
$P_{\rm KS}$ less than $0.00034$, the value found for the original data set.
For comparison, when we did the same analysis using $q(R_e)$ as the fiducial
axis ratio, we found that a fraction $f = 0.43$ of the resampled data sets
had a minimized $P_{\rm KS}$ less than $0.087$, the value for the original
data set. Thus, we may conclude that the difference in shape between young
and old galaxies is not statistically significant at $R = R_e$, but is
significant at $R = R_e/16$.

\begin{table}
\caption{Difference in axis ratio for young ($t \leq 7.5 {\rm\,Gyr}$)
and old ($ t > 7.5 {\rm\,Gyr}$) ellipticals}
\begin{tabular}{lccccl}
$R$ & $N_{\rm young}$ & $\overline{q}_{\rm young}$ &
$N_{\rm old}$ & $\overline{q}_{\rm old}$ & $P_{\rm KS}$ \\ \hline

$R_e/16$ & 23 & 0.722 & 27 & 0.835 & 0.00034 \\
$R_e/8$  & 22 & 0.721 & 30 & 0.830 & 0.0017  \\
$R_e/4$  & 29 & 0.748 & 32 & 0.829 & 0.0066  \\
$R_e/2$  & 28 & 0.747 & 32 & 0.804 & 0.077   \\
$R_e$    & 28 & 0.744 & 34 & 0.789 & 0.087   \\
$2 R_e$  & 24 & 0.763 & 22 & 0.834 & 0.037   \\
\end{tabular}
\end{table}

Terlevich \& Forbes (2000) found only a weak correlation
between apparent axis ratio and age for the elliptical
galaxies in their catalogue; however, the axis ratios
they used were measured at the effective radius, and at $R_e$,
as we have seen, the difference in shape between
young and old ellipticals is not statistically significant.

The statistical significance of the difference in shape
between young and old galaxies at small radii ($R = R_e/16$)
is not highly dependent on the choice of the dividing
time $t_0$ between old and young galaxies; $P_{\rm KS} < 0.01$
for all values of $t_0$ in the range $4.1 {\rm\,Gyr} < t_0 <
11 {\rm\,Gyr}$. A dividing time $t_0$ of several gigayears
is very much longer than the dynamical time $t_{\rm dyn}$
at $R = R_e/16$; for the galaxies in our sample,  the dynamical time
at $R = R_e/16$ is of order $t_{\rm dyn} \sim 1 {\rm\,Myr}$.

Given a sample $q_1$, $q_2$, $\ldots$, $q_N$ of apparent
axis ratios, it is possible to use kernel density estimators
to approximate the underlying distribution function $f(q)$ for
the apparent axis ratios. 
The upper panel of Figure 2 shows the kernel estimate for
the distribution of $q$ measured at small radii ($R = R_e/16$)
for young ellipticals ($t \leq 7.5 {\rm\,Gyr}$). A Gaussian
kernel was used with a bandwidth $h = 0.056$. The bandwidth
used was computed from the formula $h = 0.9 \sigma N^{-0.2}$,
where $\sigma$ is the standard deviation of the sample; this
value of $h$ minimizes the integrated mean square error for
samples which are not strongly skewed (Silverman 1986; Vio et al.
1994). The solid line
is the best fit found to the data. The dashed lines give the
80 percent confidence interval, found by bootstrap resampling
of the original data set. That is, at any value of $q$, 10 percent
of the estimates found by bootstrap resampling lie above the upper
dashed line and 10 percent lie below the lower dashed line. The
dotted lines give the 98 percent confidence interval, found by
the same bootstrap technique. The scarcity of nearly circular
isophotes ($q \gse 0.8$) is a characteristic signature of
a population of triaxial objects. However, given the small
sample size ($N = 23$) of young elliptical galaxies with
isophotal data at $R = R_e/16$, the strongest statement
we can make is that if the inner regions of young ellipticals
are randomly oriented and spheroidal (either oblate or prolate),
few of them can have
an intrinsic axis ratio $\gamma \gse 0.8$.

\begin{figure}
\centering \epsfig{figure=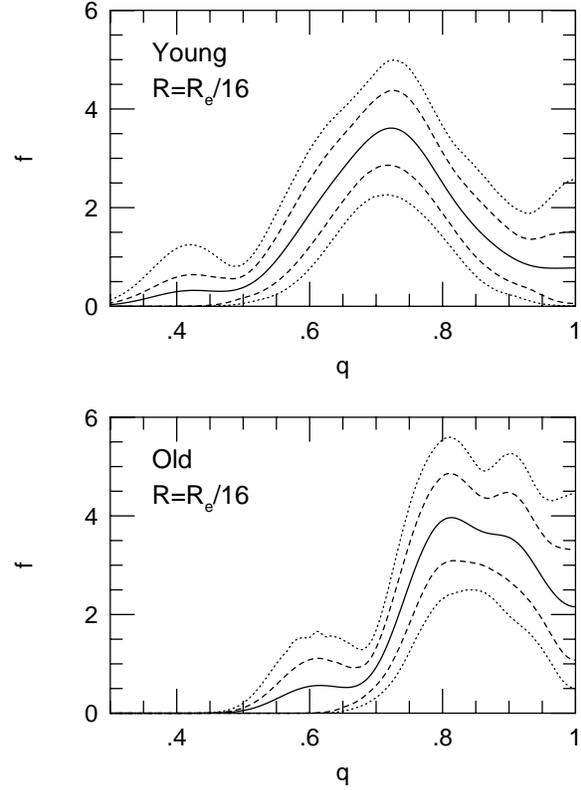, width=8.4cm}
\caption{
(Upper panel)
The distribution function for the apparent
axis ratio $q$ of young ($t \leq 7.5$ Gyr) elliptical galaxies,
as measured at $R = R_e/16$. The solid line is the best fit,
the dashed lines are the 80 percent confidence interval, and
the dotted lines are the 98 percent confidence interval.
(Bottom panel)
The same, but for the old ($t > 7.5$ Gyr) elliptical galaxies in the sample.}
\label{figure2}
\end{figure}

The lower panel of Figure 2 shows the kernel estimate for the
distribution of $q$ measured at small radii ($R = R_e/16$) for
old ellipticals ($t > 7.5 {\rm\,Gyr}$). A Gaussian kernel was
used with a bandwidth $h = 0.043$. As compared to the young ellipticals,
the old ellipticals are more likely to
have nearly circular central isophotes ($q \gse 0.8$). The observed
apparent shapes are consistent with the hypothesis that the central
regions of old ellipticals are oblate; they are also consistent
with the hypothesis that they are prolate. In sum, the apparent
shapes of young and old ellipticals are consistent with
a scenario in which the central regions of elliptical galaxies
evolve from being flattened triaxial ellipsoids to being nearly
spherical oblate spheroids; however, the data do not require such
a scenario.

\section{Core versus power-law}

The intrinsic axis ratios of an elliptical galaxy may change
with time; so may its luminosity profile. Elliptical galaxies
can be divided into two classes, based on the shape of
their luminosity profiles. Power-law ellipticals have
luminosity densities in their inner regions which are
well fitted by a pure power law all the way to the limit
of resolution. Core ellipticals, by contrast, have luminosity
densities which show a break to a shallower inner slope
(Ferrarese et al. 1994; Forbes et al. 1995; Lauer et al. 1995).
The break radius for core galaxies is generally
a few percent of the effective radius (Faber et al. 1997);
thus, for a typical core elliptical, our innermost
reference radius at $R = R_e/16$ is comparable to, or
slightly larger than, the break radius.

The discovery of the core/power-law dichotomy has led
to speculation about its cause. In the scenario of
Faber et al. (1997), power-law ellipticals form in
gas-rich mergers. In such a merger, dissipation and
angular momentum transfer permits the gas to fall
toward the centre, and naturally give rise to a
steep power-law profile in the merger remnant
(Barnes \& Hernquist 1991, 1996; Mihos \& Hernquist
1994, 1996). We might expect the gaseous nature of the
power-law galaxies' formation to be accompanied
by a burst of star formation and the formation of
a stellar disc (de Jong \& Davies 1997). The origin
of core ellipticals is more problematic. Faber et al.
(1997) suggest that core galaxies form in largely
dissipationless mergers; a low-density core is
`scoured out' by the black hole binary which forms
by orbital decay of the black holes in the progenitor
galaxies (Eisuzaki et al. 1991; Makino \& Eisuzaki 1996).
The central black hole which results from the coalescence
of the binary maintains the central low density of the
core by tidally disrupting any high-density satellite
galaxies subsequently accreted by the core elliptical
(Faber et al. 1997; Merritt \& Cruz 2001).

To test the correlation among age, profile type, and isophote axis ratio,
we have determined the profile type (core or power-law) for as many
of the 74 ellipticals in our galaxy sample as possible (either
from the literature or deriving them ourselves). The resulting
profile types are given in Table 1. In assigning profile types,
we followed the definition of Faber et al. (1997) that a
core galaxy has a surface density profile with a logarithmic
slope $d \log I / d \log R > -0.3$ inside its break radius.
In the case of NGC 7626, HST WFPC2 imaging yields a core profile
at $V$ and $I$ (Carollo et al. 1997), but NICMOS imaging
yields a power-law profile at $1.6 \mu{\rm m}$ (Quillen et al. 2000).
We assign NGC 7626 a core profile, for consistency with
other systems for which only $V$ or $I$ images are available.

Of the 74 ellipticals in our sample, 53 had HST images which
permitted classification of their profile type. The 22
power-law galaxies have a mean and standard deviation for
their ages of $t = 6.9 \pm 3.5 {\rm\,Gyr}$. For the
29 core galaxies, the mean and standard deviation are
$t = 8.6 \pm 3.3 {\rm\,Gyr}$. Figure 3 shows the
distribution function for the ages of the power-law
galaxies (top panel) and the core galaxies (bottom
panel). A KS test comparing the distribution of
ages for the two different types of galaxy yields
$P_{\rm KS} = 0.017$.

\begin{figure}
\centering \epsfig{figure=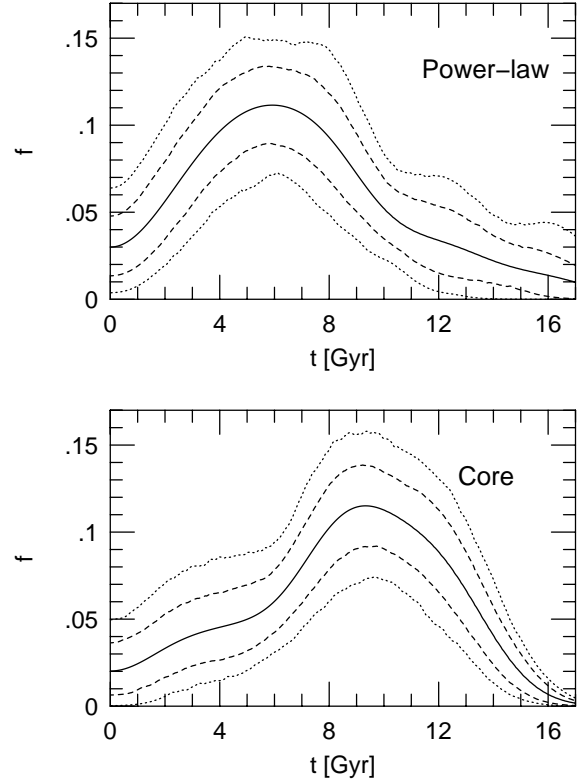, width=8.4cm}
\caption{
(Upper panel) The distribution function for the ages of
power-law galaxies, as found by a kernel
density estimator. The solid line is the best fit, the
dashed lines are the 80 percent confidence interval, and
the dotted lines are the 98 percent confidence interval.
(Bottom panel) The same information for the core galaxies
in the sample.}
\label{figure3}
\end{figure}

{\it The power-law galaxies are significantly younger, on average,
than the core galaxies}. Of the power-law galaxies, 13 fall
into the `young' category ($t \leq 7.5 {\rm\,Gyr}$) and
9 into the `old' category. Of the core galaxies, by contrast,
only 8 are young and 23 are old. 
Since profile type correlates
with age, and age correlates with axis ratio (in
the central regions of a galaxy), it is not surprising that
profile type correlates with axis ratio. In Figure 1, the
core galaxies (symbolised by squares) and the power-law galaxies
(symbolised by triangles) lie in different regions of the $q - t$
plane. Particularly when $q$ is measured at $R = R_e/16$, it
is seen that power-law galaxies are younger and flatter;
core galaxies are older and rounder. Table 3 gives the
numbers $N_{\rm power}$ and $N_{\rm core}$ of power-law
and core ellipticals in our sample at each reference radius,
as well as the mean axis ratios $\overline{q}_{\rm power}$
and $\overline{q}_{\rm core}$ and the KS probability
measuring the difference in the shape distribution for
power-law and core ellipticals. Note that at most radii,
the difference in shape between power-law and core ellipticals
is comparable in significance to the difference in
shape between young and old ellipticals. This difference
in shape between core and power-law ellipticals was 
indirectly uncovered by Tremblay \& Merritt (1996), who
found that ellipticals brighter than $M_B = -20$ were
rounder in projection than fainter ellipticals. Since
ellipticals brighter than $M_B = -20$ are predominantly
core galaxies while those fainter are predominantly power-law
galaxies, the dependence of shape on luminosity is a
reflection of the dependence of shape upon profile type.

\begin{table}
\caption{Difference in axis ratio for power-law and core ellipticals}
\begin{tabular}{lccccl}
$R$ & $N_{\rm power}$ & $\overline{q}_{\rm power}$ &
$N_{\rm core}$ & $\overline{q}_{\rm core}$ & $P_{\rm KS}$ \\ \hline

$R_e/16$ & 21 & 0.705 & 28 & 0.843 & 0.00041\\
$R_e/8$  & 21 & 0.721 & 29 & 0.830 & 0.0033 \\
$R_e/4$  & 21 & 0.723 & 29 & 0.830 & 0.012  \\
$R_e/2$  & 21 & 0.726 & 25 & 0.822 & 0.027  \\
$R_e$    & 21 & 0.720 & 26 & 0.808 & 0.020  \\
$2 R_e$  & 16 & 0.762 & 20 & 0.828 & 0.35   \\
\end{tabular}
\end{table}

The upper panel of Figure 4 shows the kernel estimate for
the distribution of $q$ at small radii ($R = R_e/16$)
for the power-law ellipticals in our sample. A Gaussian
kernel was used with a bandwidth $h=0.056$. Note,
in the best-fitting estimate (given by the solid line),
the scarcity of power-law galaxies with nearly 
circular isophotes. Of the 21 power-law ellipticals
with isophotal information at $R_e/16$, only one
(NGC 4339) has $q > 0.82$. This scarcity of round
isophotes is highly implausible in a population of
randomly oriented oblate spheroids. For such a population,
the distribution function of the intrinsic axis ratio $\gamma$
that minimizes the number of nearly circular isophotes
is a delta function at $\gamma = 0$ (reflecting a population
of infinitesimally thin discs). This distribution for $\gamma$
produces a uniform distribution for $q$ over the range $0 \leq q
\leq 1$. Thus, for a population of infinitesimally thin circular discs,
which {\it minimizes} the number of nearly circular isophotes given
the oblate hypothesis, the probability of seeing
0 or 1 galaxies out of 21 with $q > 0.82$
is only $P = (0.82)^{21} + 21 (0.82)^{20}(0.18) = 0.086$.
For a more plausible distribution of $\gamma$ (that is,
one that would not overestimate the number of galaxies
with small $q$), the probability $P$ of finding so
few nearly circular isophotes would be smaller still.

\begin{figure}
\centering \epsfig{figure=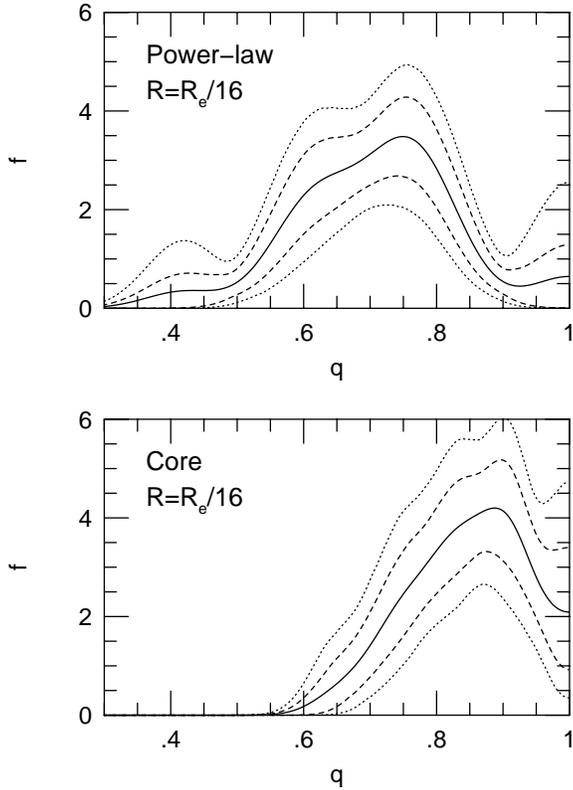, width=8.4cm}
\caption{(Upper panel) The distribution function for the apparent
axis ratio $q$ of power-law ellipticals, as measured at $R = R_e/16$.
The solid line is the best fit, the dashed lines are the 80 percent
confidence level, and the dotted lines are the 98 percent confidence
level.
(Bottom panel) The same, but for the core ellipticals in the sample.}
\label{figure4}
\end{figure}

The distribution of $q$ for core ellipticals, as
measured at $R = R_e/16$, is displayed in the lower panel of
Figure 4. A Gaussian kernel was used, with $h = 0.039$.
For the sample of core galaxies, there
is no scarcity of nearly circular isophotes. Of
the 28 core ellipticals with isophotal information
at $R_e/16$, 18 have $q > 0.82$. Neither the oblate
or prolate hypothesis can be rejected. If the inner regions
of core ellipticals are oblate, the mean intrinsic axis
ratio is $\overline{\gamma}_o = 0.75$;
if the inner regions are prolate, the mean intrinsic
axis ratio is $\overline{\gamma}_p = 0.78$.

In summary, the distribution of axis ratios for
power-law ellipticals, as measured at $R = R_e/16$,
is inconsistent with the hypothesis that their
central regions are randomly oriented oblate spheroids.
The distribution of axis ratios for core
ellipticals, at the same reference radius, is consistent
with the hypothesis that they are randomly oriented
and spheroidal (either oblate or prolate).

\section{Implications}

We have shown that elliptical galaxies with estimated assembly ages
$t > 7.5 {\rm\,Gyr}$ are rounder in projection than younger
ellipticals; the statistical significance of the shape
difference is greater at smaller radii. Similarly, we
have shown that core elliptical galaxies are rounder in projection
than power-law ellipticals; the significance of the
shape difference is likewise greater at smaller radii.
It is no coincidence that the solid lines in the two
panels of Figure 3, giving the age distribution for power-law
and core galaxies, intersect at an age $t \approx 7.5 {\rm\,Gyr}$.
Our sample of old galaxies, with $t > 7.5 {\rm\,Gyr}$, consists
primarily of core ellipticals, while our sample of young galaxies,
with $t \leq 7.5 {\rm\,Gyr}$, consists mainly of power-law galaxies.

Our results are consistent with the Faber et al. (1997)
scenario, in which power-law galaxies form in gas-rich
mergers and core ellipticals form in dissipationless
mergers, with shallow density cores created and maintained
by the influence of central black holes. The core elliptical
galaxies have relatively old stellar ages, in this
scenario, since their most recent major merger was
dissipationless, without an accompanying burst of star
formation. Since the core scoured out by a central
black hole tends to be nearly spherical (Merritt 2000), it would
also explain why the central isophotes of core ellipticals
tend to be nearly spherical. By contrast, the gas-rich
mergers which form power-law galaxies are accompanied
by a burst of star formation, and tend result in
the formation of a central stellar disc (de Jong \& Davies
1997). Thus, the age $t$ of a power-law galaxy reflects
the time since the merger in which it was created; the
flattened central isophotes of power-law galaxies reflect
the presence of the disc which formed in the merger.

One question that arises, though, is why there are so
few old power-law galaxies. There was no shortage of
gas-rich mergers at $t > 7.5 {\rm\,Gyr}$; indeed, the
ratio of gas-rich to gas-poor mergers tends to decrease
with time. One way in which old power-law galaxies are
destroyed is through mergers. If a power-law galaxy
undergoes a dissipational merger, the resulting galaxy
will still have a power-law profile, but will have a
younger age $t$, thanks to the star formation which
accompanies a dissipational merger. If, by contrast,
a power-law galaxy undergoes a dissipationless merger
with a large core elliptical, the resulting galaxy
will have a core profile, as the central density cusp
of the power-law galaxy is disrupted by the central
black hole of the core elliptical (Merritt \& Cruz 2001).
Thus, mergers tend to convert old power-law galaxies
(where `old' is a description of the age of the
central stellar population) into young power-law galaxies
if the merger is gas-rich, or into old core galaxies
if the merger is gas-poor.

At all events, the observed correlation among luminosity
profile type, axis ratio, and age of stellar population
provides a useful constraint for all future studies of
the origin of the core/power-law dichotomy.

\section*{Acknowledgments}

We would like to thank M. Merrifield for acting as matchmaker.
I. Jorgensen and A. Quillen kindly provided data in electronic form.
D. Merritt, T. Lauer, D. Richstone, and the anonymous referee made
useful comments. This research has made use of the NASA/IPAC Extragalactic
Database (NED) which is operated by the Jet Propulsion Laboratory,
California Institute of Technology, under contract with the National
Aeronautics and Space Administration.

\label{lastpage}

\begin{thebibliography}{}

\bibitem[\protect\citename{Barnes }1992]{ba92}
Barnes, J. 1992, ApJ, 393, 484

\bibitem[\protect\citename{Barnes \& Hernquist }1991]{bh91}
Barnes, J. E., \& Hernquist, L. 1991, ApJ, 370, L65

\bibitem[\protect\citename{Barnes \& Hernquist }1996]{bh96}
Barnes, J. E., \& Hernquist, L. 1996, ApJ, 471, 115

\bibitem[\protect\citename{Bettoni et al. }1997]{bf97}
Bettoni, D., Fasano, G., Kjaergaard, P., \& Moles, M. 1997,
in Arnaboldi, M., Da Costa, G. S., \& Saha, P. eds, The
Second Stromlo Symposium: The Nature of Elliptical Galaxies.
ASP, San Francisco, p. 71

\bibitem[\protect\citename{Caon et al. }1990]{cc90}
Caon, N., Capaccioli, M., \& Rampazzo, R. 1990, A\&AS, 86, 429

\bibitem[\protect\citename{Caon et al. }1994]{cc94}
Caon, N., Capaccioli, M., \& D'Onofrio, M. 1994, A\&AS, 106, 199

\bibitem[\protect\citename{Capaccioli et al. }1988]{cp88}
Capaccioli, M., Piotto, G., \& Rampazzo, R. 1988, AJ, 96, 487

\bibitem[\protect\citename{Carollo et al. }1997]{cf97}
Carollo, C. M., Franx, M., Illingworth, G. D., \& Forbes, D. A. 1997,
ApJ, 481, 710

\bibitem[\protect\citename{de Jong \& Davies }1997]{dd97}
de Jong, R. S., \& Davies, R. L. 1997, MNRAS, 285, L1

\bibitem[\protect\citename{de Juan et al. }1994]{dc94}
de Juan, L., Colina, L., Perez-Fournon, I. 1994, ApJS, 91, 507

\bibitem[\protect\citename{Ebisuzaki et al. }1991]{em91}
Ebisuzaki, T., Makino, J., \& Okumura, S. K. 1991, Nature, 354, 212

\bibitem[\protect\citename{Faber et al. }1989]{fw89}
Faber, S. M., Wegner, G., Burstein, D., Davies, R. L., Dressler, A.,
Lynden-Bell, D., \& Terlevich, R. J. 1989, ApJS, 69, 763

\bibitem[\protect\citename{Faber et al. }1997]{ft97}
Faber, S. M., et al. 1997, AJ, 114, 1771

\bibitem[\protect\citename{Ferrarese et al. }1994]{fv94}
Ferrarese, L., van den Bosch, F. C., Ford, H. C., Jaffe, W.,
O'Connell, R. W. 1994, AJ, 108, 1598

\bibitem[\protect\citename{Forbes et al. }1995]{ff95}
Forbes, D. A., Franx, M., \& Illingworth, G. D. 1995, AJ, 109, 1988

\bibitem[\protect\citename{Franx et al. }1989]{fi89}
Franx, M., Illingworth, G., \& Heckman, T. 1989, AJ, 98, 538

\bibitem[\protect\citename{Gerhard \& Binney }1985]{gb85}
Gerhard, O. E., \& Binney, J. 1985, MNRAS, 216, 467

\bibitem[\protect\citename{Goudfrooij et al. }1994]{gh94}
Goudfrooij, P., Hansen, L., Jorgensen, H. E., Norgaard-Nielsen,
H. U., de Jong, T., \& van den Hoek, L. B. 1994, A\&AS, 104, 179

\bibitem[\protect\citename{Hubble }1926]{hu26}
Hubble, E. 1926, ApJ, 64, 321

\bibitem[\protect\citename{Jedrzejewski }1987]{je87}
Jedrzejewski, R. 1987, MNRAS, 226, 747

\bibitem[\protect\citename{Jedrzejewski et al. }1987]{jd87}
Jedrzejewski, R. I., Davies, R. L., \& Illingworth, G. D. 1987,
AJ, 94, 1508

\bibitem[\protect\citename{Jorgensen et al. }1992]{jf92}
Jorgensen, I., Franx, M., \& Kjaergaard, P. 1992,
A\&AS, 95, 489

\bibitem[\protect\citename{Kormendy \& Richstone }1995]{kr95}
Kormendy, J., \& Richstone, D. 1995, ARA\&A, 33, 581

\bibitem[\protect\citename{Lauer }1985]{la85}
Lauer, T. R. 1985, ApJS, 57, 473

\bibitem[\protect\citename{Lauer et al. }1995]{la95}
Lauer, T. R. et al. 1995, AJ, 110, 2622

\bibitem[\protect\citename{Lauer et al. }1998]{lf98}
Lauer, T. R., Faber, S. M., Ajhar, E. A., Grillmair, C. J., \&
Scowen, P. A. 1998, AJ, 116, 2263

\bibitem[\protect\citename{Lauer et al. }2000]{la00}
Lauer, T. R., et al. 2000, in preparation

\bibitem[\protect\citename{Magorrian et al. }1998]{ma98}
Magorrian, J., et al. 1998, AJ, 115, 2285

\bibitem[\protect\citename{Makino \& Ebisuzaki }1996]{mk96}
Makino, J., \& Ebisuzaki, T. 1996, ApJ, 465, 527

\bibitem[\protect\citename{Mehlert et al. }2000]{ms00}
Mehlert, D., Saglia, R. P., Bender, R., \&
Wegner, G. 2000, A\&AS, 141, 449

\bibitem[\protect\citename{Merritt }2000]{me00}
Merritt, D. 2000, preprint (astro-ph/9910546)

\bibitem[\protect\citename{Merritt }2001]{me01}
Merritt, D. 2001, ApJ, submitted (astro-ph/00012264)

\bibitem[\protect\citename{Merritt \& Cruz}2001]{mc01}
Merritt, D., \& Cruz, F. 2001, ApJ, submitted (astro-ph/0101194)

\bibitem[\protect\citename{Merritt \& Quinlan }1998]{mq98}
Merritt, D., \& Quinlan, G. 1998, ApJ, 498, 625

\bibitem[\protect\citename{Mihos \& Hernquist }1994]{mh94}
Mihos, J. C., \& Hernquist, L. 1994, ApJ, 431, L9

\bibitem[\protect\citename{Mihos \& Hernquist }1996]{mh96}
Mihos, J. C., \& Hernquist, L. 1996, ApJ, 464, 641

\bibitem[\protect\citename{Norman et al. }1985]{nm85}
Norman, C. A., May, A., \& van Albada, T. S. 1985, ApJ, 296, 20

\bibitem[\protect\citename{Peletier }1993]{pe93}
Peletier, R. F. 1993, A\&A, 271, 51

\bibitem[\protect\citename{Peletier et al. }1990]{pd90}
Peletier, R. F., Davies, R. L., Illingworth, G. D., Davis, L. E.,
\& Cawson, M. 1990, AJ, 100, 1091

\bibitem[\protect\citename{Postman \& Lauer }1995]{pl95}
Postman, M., \& Lauer, T. R. 1995, ApJ, 440, 28

\bibitem[\protect\citename{Quillen et al. }2000]{qb00}
Quillen, A., Bower, G. A., \& Stritzinger, M. 2000,
ApJS, submitted (astro-ph/9907021)

\bibitem[\protect\citename{Quinlan \& Hernquist }1997]{qh97}
Quinlan, G. D., \& Hernquist, L. 1997, NewA, 2, 533

\bibitem[\protect\citename{Ryden }1996]{ry96}
Ryden, B. S. 1996, ApJ, 461, 146

\bibitem[\protect\citename{Schweizer et al. }1989]{sv89}
Schweizer, F., van Gorkom, J. H., \& Seitzer, P. 1989,
ApJ, 338, 770

\bibitem[\protect\citename{Silverman }1986]{si86}
Silverman, B. W. 1986, Density Estimation for Statistics
and Data Analysis (New York: Chapman \& Hall)

\bibitem[\protect\citename{Sparks et al. }1991]{sw91}
Sparks, W. B., Wall, J. V., Jorden, P. R., Thorne, D. J., \&
van Breda, I. 1991, ApJS, 76, 471

\bibitem[\protect\citename{Springel }2000]{sp00}
Springel, V. 2000, MNRAS, 312, 859

\bibitem[\protect\citename{Terlevich \& Forbes }2000]{tf00}
Terlevich, A. I., \& Forbes, D. A., 2000, MNRAS, submitted

\bibitem[\protect\citename{Tremblay \& Merritt }1996]{tm96}
Tremblay, B., \& Merritt, D. 1996, AJ, 111, 2243

\bibitem[\protect\citename{Valluri \& Merritt }1998]{vm98}
Valluri, M., \&  Merritt, D. 1998, ApJ, 506, 686

\bibitem[\protect\citename{van den Bosch et al. }1994]{vd94}
van den Bosch, F. C., Ferrarese, L., Jaffe, W., Ford, H. C., \&
O'Connell, R. W., 1994, AJ, 108, 1579

\bibitem[\protect\citename{Verdoes Kleijn et al. }1999]{vk99}
Verdoes Kleijn, G. A., Baum, S., de Zeeuw, P. T., \& O'Dea, C. P.
1999, AJ, 118, 2592

\bibitem[\protect\citename{Vio et al. }1994]{vf94}
Vio, R., Fasano, G., Lazzarin, M., \& Lessi, O. 1994, A\&A, 289, 640

\end{thebibliography}
\end{document}